\documentstyle[12pt,aasms]{article}
\slugcomment{Submitted to \it{Publications of the Astronomical Society of the
Pacific}}
\begin{document}
\title{A Generalized $K$ correction for Type Ia Supernovae: \\
Comparing
$R$-band Photometry Beyond $z=0.2$ \\
with $B$, $V$, and $R$-band Nearby Photometry}
\author{ Alex Kim\altaffilmark{1}
and Ariel Goobar\altaffilmark{2}
and Saul Perlmutter \altaffilmark{3}
}
\altaffiltext{1}{Center for Particle Astrophysics and
Lawrence Berkeley Laboratory, 50-232,
Berkeley, CA  94720; agkim@lbl.gov}
\altaffiltext{2}{Physics Department, Stockholm University, Box 6730,
S-113 85 Stockholm, Sweden; ariel@physto.se}
\altaffiltext{3}{
Lawrence Berkeley Laboratory, 50-232, and
Center for Particle Astrophysics, University of California, 
Berkeley, CA  94720; saul@lbl.gov}

\begin{abstract}
Photometric measurements show that as a group nearby type Ia supernovae 
follow similar lightcurves and reach similar peak magnitudes
(Branch \& Tammann 1992)\nocite{br:araa}.
Thus, these supernovae can serve as standard candles or calibrated
candles at cosmological
distances.\nocite{br:araa}\nocite{le:thesis}
Magnitudes of local and distant supernovae, both in the same filter band,
are compared
using a $K$ correction
to account for the different spectral
regions incident on that filter.
A generalized
approach compares magnitudes in different bands for the nearby and distant
supernova, bands that are selected to give
sensitivity in corresponding regions of the redshifted and unredshifted spectra.
Thus at a redshift
of $z \approx 0.5$, local $B$ magnitudes are compared with distant
$R$ magnitudes.
We compute these generalized $K$ corrections over a range of
redshifts and bandpass pairs and discuss their advantages 
over the traditional single-band $K$ correction.  In particular, errors
near maximum light
can be kept below 0.05 mag out to at least
$z=0.6$, whereas the traditional $K$ correction
is difficult to use beyond $z > 0.2$.
\end{abstract}

\section{Introduction}

Assuming no evolution and a homogeneous set of type Ia supernovae, 
a perfect and complete set of type Ia supernova spectra
can be used to calculate the apparent magnitudes
at any redshift and epoch,
given the transmission functions and the zeropoints of the magnitude
system.
Unfortunately, this is not feasible due to several problems: (1) The
available supernova spectra often have insufficient
wavelength coverage to calculate broadband photometry and insufficient
time coverage to track the quickly evolving supernova;
(2) Many of the available spectra do not have the signal-to-noise
to calculate precise magnitudes;
(3) Spectral miscalibrations can lead to large errors in magnitude
determinations;
(4) The filter transmission function and detector response function are not
perfectly known.
More reliable magnitude calculations can be made using spectra and photometry
together, photometry being less sensitive to the above problems than spectra.
In this paper, 
we calculate and discuss the errors for a
generalized $K$ correction, an example of such a technique, with a
preliminary analysis using three ``normal''
type Ia supernova.  (One ``peculiar'' was also examined for comparison
purposes.)
These $K$ corrections are particularly important for
use with supernovae at $z > 0.2$, which are now being discovered
in systematic searching (e.g., Perlmutter et al. 1994, 1995).
\nocite{sn1992bi}\nocite{sne1994}

\section{A Generalized $K$ Correction}

The standard $K$ correction, $K_x$, is used to calculate the
apparent magnitude in some ``$x$'' filter band of an object at redshift $z$
according to the equation\nocite{ok:kcorr}: 
$m_x(z,t)=M_x(t)+\mu(z)+K_x(z,t)$, where $\mu$ is the distance modulus (based on
luminosity distance) and  $M_x$ is the absolute $x$ magnitude
(we omit explicit time dependence in
subsequent equations).
The $K$ correction relates nearby and distant magnitudes measured
with the same filter:
\begin{equation}
  K_x = 2.5 \log(1+z) +
  2.5 \log
    \left( 
    \frac
	{\int F(\lambda)S_x(\lambda)d\lambda}
	{\int F(\lambda/(1+z))S_x(\lambda)d\lambda}
    \right)
\label{Ki}
\end{equation}
where $F(\lambda)$ is the spectral energy distribution at the source (in
this case the supernova), and
$S_x(\lambda)$ is the $x$'th filter transmission (Oke \& Sandage 1968).

\vspace{.1in}
  We generalize this expression to handle different filters, adding
  a term that accounts for the differences in the zeropoints of the
  magnitude system:
\begin{eqnarray}
  K_{xy} & = &  -2.5 \log
    \left(
    \frac
       {\int {\cal Z}(\lambda)S_x(\lambda)d\lambda}
       {\int {\cal Z}(\lambda)S_y(\lambda)d\lambda}
    \right)
    +2.5 \log(1+z) 
    +2.5 \log
    \left( 
    \frac
	{\int F(\lambda)S_x(\lambda)d\lambda}
	{\int F(\lambda/(1+z))S_y(\lambda))d\lambda}
    \right)\nonumber \\
    & = & -2.5 \log
    \left(
    \frac
       {\int {\cal Z}(\lambda)S_x(\lambda)d\lambda}
       {\int {\cal Z}(\lambda)S_y(\lambda)d\lambda}
    \right)
    +2.5 \log
    \left( 
    \frac
	{\int F(\lambda)S_x(\lambda)d\lambda}
	{\int F(\lambda')S_y(\lambda'(1+z))d\lambda'}
    \right)
\label{Kij}
\end{eqnarray}
where 
${\cal Z}(\lambda)$ is an idealized stellar SED 
for which $U=B=V=R=I=0$ in the photometric system being used.
$K_{xy}$ is thus defined so that $m_y=M_x+\mu+K_{xy}$.  If
$S_x \equiv S_y$, the first term drops out and this reduces to 
the standard $K$ correction of Equation~\ref{Ki}.

\begin{figure}
  \plotone{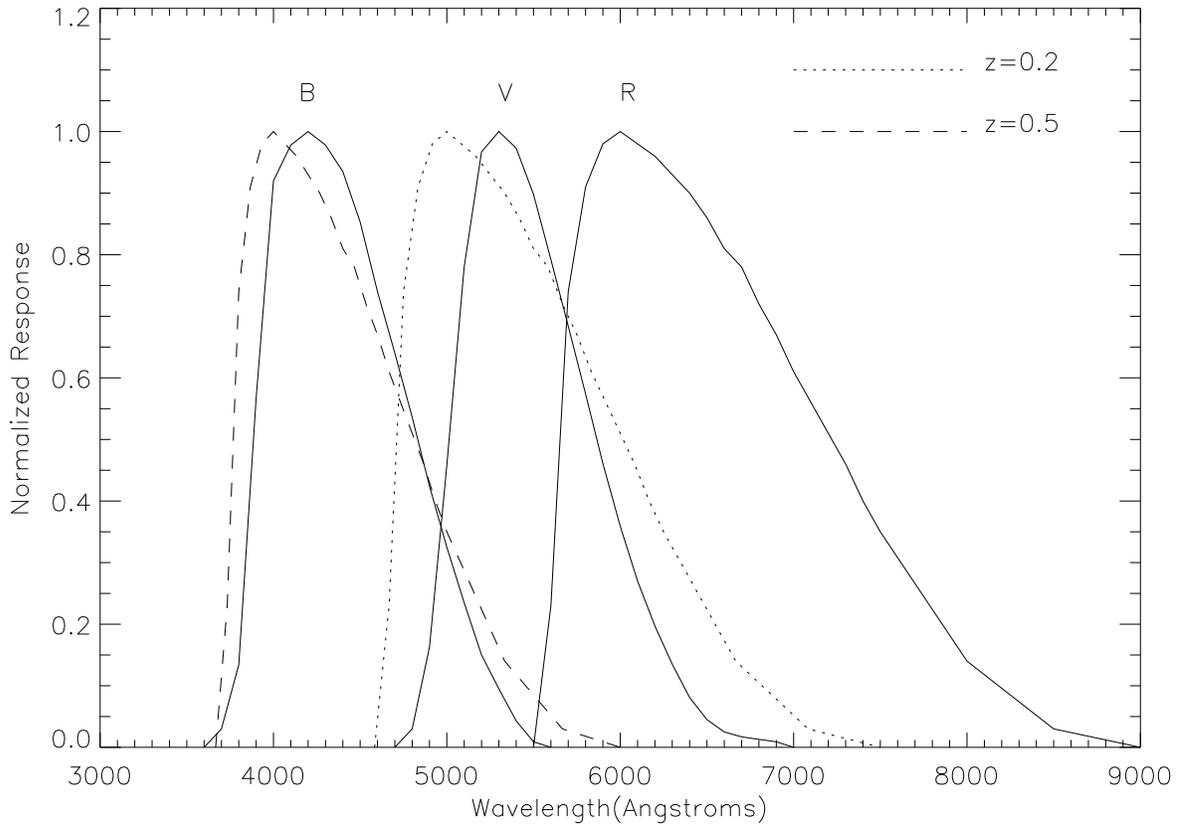}
  \caption[figurecaption]{
  Bessell's representations of the Johnson-Cousins $B$, $V$, and $R$ passbands.
  The dotted lines represent the blue-shifted $R$ filter for $z=0.2$
  and $z=0.5$.  The $R$ filter approximately matches the $V$ and $B$ filters at these
  redshifts.}
  \label{filters}
\end{figure}

\vspace{.1in}
The second line of Equation~\ref{Kij} is a change of variables,
$\lambda'= \lambda/(1+z)$,
that makes it easier to understand the $K_{xy}$ correction in the
case $S_y(\lambda (1+z)) = S_x(\lambda)$, a situation
approximated by the dashed lines
in Figure~\ref{filters}.  If the ``blueshifted''
$y$'th filter matches the $x$'th filter function the second
term in this equation drops out, and one is left with the term
accounting for the difference in zeropoints of the filters (this
difference is the ``color zeropoint'').
In this case, spectral dependence on the correction is eliminated.
Note that this cross-filter
approach has previously been used for galaxy $K$ corrections
(e.g. Gunn 1978).\nocite{gu:kcorr}

\section{$K_{xy}$ Calculation}

We calculate generalized $K$ corrections using Equation~\ref{Kij}
with Bessell's (1990) \nocite{be:ubvri} color zeropoints and
realizations of the Johnson-Cousins UBVRI
filter system (Figure~\ref{filters}).
The color zeropoints are expected to match real photometric color 
zeropoints to better than
0.01 mag (Hamuy et al. 1992\nocite{ha:so} quotes $\leq 0.009$, Bessell 
(private communication)
quotes $\leq 0.005$).
Only supernovae with non-peculiar spectra and no evidence for reddening
are used, as described by Vaughan et al. (1995).\nocite{va:inp}
Our full sample is presented in Table~1 and
contains 29 spectra from epochs
$-14<t_{max}^{B}<76$
days after blue maximum for supernovae SN1981B, SN1990N, and SN1992A.
The SN1981B data is from Branch et al. (1983)\nocite{br:sn1981b}, SN1990N
data is described in
Leibundgut et al. (1991), and SN1992A is in
Suntzeff et al. (1995)\nocite{su:prep},
and Kirshner et al. (1993)\nocite{ki:sn1992a}.
The SN1981B spectra labeled by epoch (0) is a composite of
four
spectra from March 6-9 (Branch et al. 1983).
The SN1992A HST spectra from epoch 5 with a spectral range
of 1650-4800 {\AA} has been augmented
by the CTIO spectra from epoch 6 as described in Kirshner et al. (1993);
it is labeled epoch (5).
The $K$ corrections were not calculated for cases in which the spectra did not 
cover at least $99\%$ of the effective acceptance of
the passband and are labeled with ellipsis.

Tables~2,~3, and~4 have $K_{xy}$ corrections for $x=B,V,R$, $y=R$ and
redshifts spanning from 0 to 0.7 in increments of 0.025. 
Epochs are given in the supernova rest-frame.  Note that $K_{RR}$
is just the standard $R$ band $K$ correction. 
Each column of data is for a single supernova spectrum; the three tables have
different number of columns because the number of spectra
with sufficient wavelength coverage to calculate each correction varies.
In particular,
Table~4 has much fewer entries because there are only a few spectra
covering the needed wavelength range to calculate $K_{RR}$.  See Hamuy et al.
for tables of $K_{BB}$ and $K_{VV}$; these $K$ corrections have poor spectral
coverage at $z > 0.1$ for $K_{BB}$ and at $z > 0.3$ for $K_{VV}$.

%
%

\section{Error Estimates and Determination of Optimal Filter Pair}

We consider the contributions of
the following sources of error in the $K$ correction:
numerical integration error, spectral measurement and calibration error,
intrinsic supernova-to-supernova dispersion, 
instrumental effects, and zeropoint uncertainty.

The numerical integration is accurate to 0.001 mag and we are able to
reproduce the standard $K$ corrections in $B$ and in $V$ of Hamuy et al.
\nocite{ha:kcorr} to that accuracy.
A larger source of
uncertainty comes from the noise and calibration error of the spectra
themselves.   Lacking prior spectral error information,
we test each spectrum's error properties
by calculating, for $z=0$, $B-V$ colors on the subset
of 24 spectra with sufficient coverage.\nocite{su:sn1992}\nocite{le:sup}
\nocite{le:sn1990n}Their difference from the photometrically
observed colors form a Gaussian distribution with a sigma of 0.04 mag.
These $B-V$ colors compare two spectral regions that have
little overlap, while $K$ corrections that compare overlapping regions
are less sensitive to large scale
miscalibrations and therefore should have smaller
error.

\begin{figure}
  \plotone{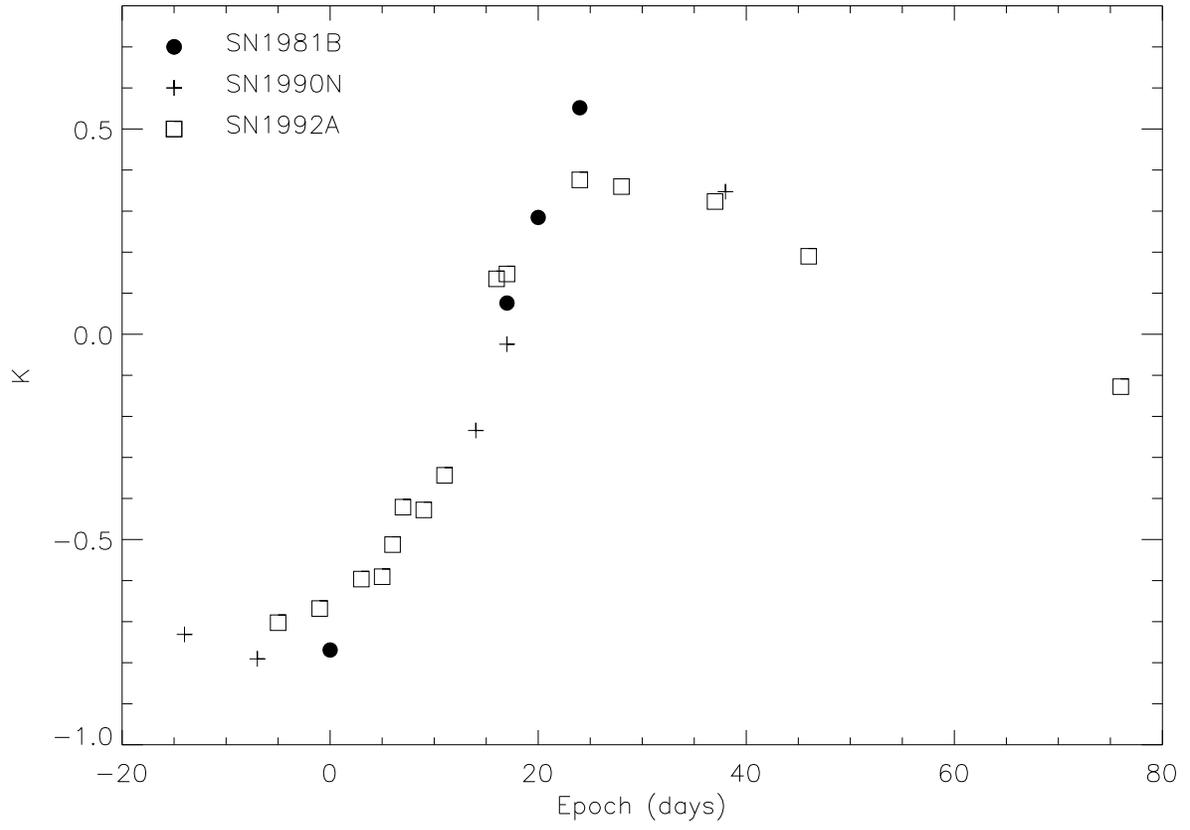}
  \caption[figurecaption]{
  $K_{VR}(z=0.5)$ as a function of epoch for SN1981B, SN1990N, and SN1992A.}
  \label{kvr5}
\end{figure}

\begin{figure}
  \plotone{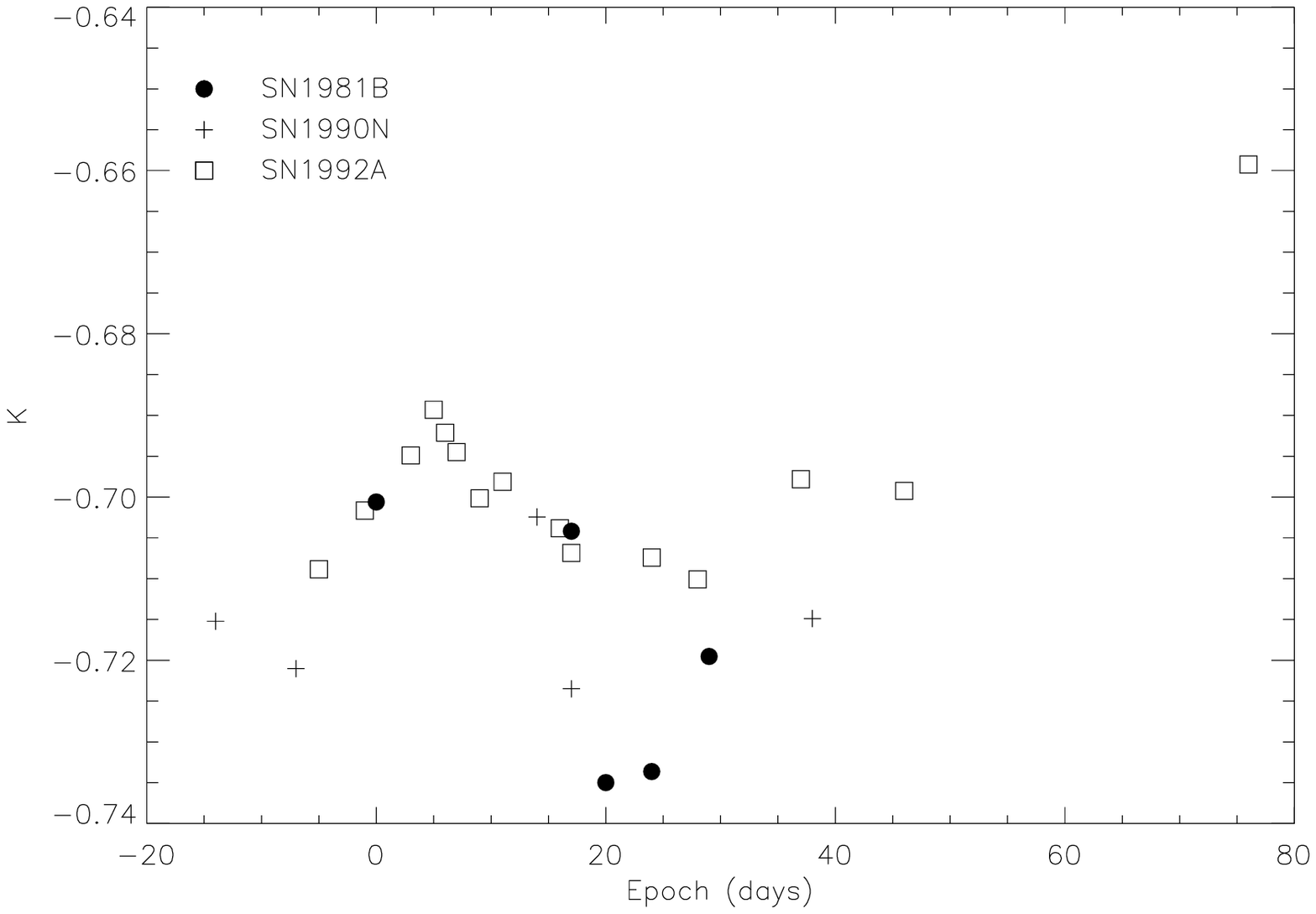}
  \caption[figurecaption]{
  $K_{BR}(z=0.5)$ as a function of epoch for SN1981B, SN1990N, and SN1992A.
  The same data as in Figure~5(b) but on a blown up scale.}
  \label{kbr5}
\end{figure}

The rapid but smooth evolution of supernova spectra should make
$K$ corrections a smooth function on the scale of a few days.
However, the data shows scatter from measurement and calibration
error in the spectra.
The good temporal sampling of SN1992A allows us to make $K$ correction
error estimates
based on this scatter.  We illustrate by considering the specific
examples of $K_{BR}$ and $K_{VR}$ at $z=0.5$, shown in
Figures~\ref{kvr5} and~\ref{kbr5}.
(Unfortunately, there are insufficient data
points to similarly consider $K_{RR}$.)
We first study the subset of SN1992A $K$ corrections calculated from
spectra measured at a single telescope,
the CTIO 1.0-m.  This includes all SN1992 spectra
except the ones at epochs 5, 6, 9, 17, 46, and 76.
We estimate the root-mean-square scatter of the data from a smooth curve to be
$< 0.002$ for $K_{BR}$ and $< 0.02$ for $K_{VR}$;
we take this to be the bound on the effects of spectral measurement errors.
Considering the full sample of SN1992A $K$ corrections
from all the telescopes, we find the
range (not the root-mean-square)
in scatter to be $\sim$0.004 for $K_{BR}$ and $\sim$0.1
for $K_{VR}$; we take these to be the bounds on
systematic instrumental error.

Having studied errors involved in the $K$ corrections of a single type Ia
supernova, we now
consider the uncertainties involved in constructing a single $K$ correction for
this entire class of supernovae.  We do this by examining systematic differences
between the three supernovae in our sample.
Again consider $K$ corrections for $z=0.5$ plotted in Figures~\ref{kvr5}
and~\ref{kbr5}.  
The scatter at a given epoch in $K_{BR}$ and $K_{VR}$
is dominated by intrinsic supernova-to-supernova differences. 
These differences are understood as being due to
the observed variance in these particular supernovae's color evolution,
particularly around 20 days after maximum when the supernovae
have quickly reached their reddest color.
The range of scatter is $\sim$0.015 mag for $K_{BR}$ and
$\sim$0.2 mag for $K_{VR}$; for epochs before day 17, the range narrows to
$< 0.002$ for $K_{BR}$ and $\sim$0.1 mag for $K_{VR}$.

As a preliminary test of the variation in $K$ correction for type Ia 
supernovae that
are not ``normal,'' the $K$ corrections of the ``peculiar'' type Ia
SN1991T were also calculated\nocite{ph:sn1991t}\nocite{fo:sn1991t}
(Ford et al. (1993) and Phillips et al. (1992) describe this supernova).
Their scatter with respect to the normal supernovae's $K$ corrections
was within the intrinsic supernova-to-supernova
dispersion discussed above,  even for epochs $< 14$,
when the SN1991T spectra least resembled
``normal'' type Ia spectra.
This suggests that broadband averaging smooths over
intrinsic differences in supernova spectra and that $K$ corrections may be
robust enough to use for the entire class of type Ia supernovae,
although more examples of peculiar type Ia supernovae (e.g. SN1991bg)
will be needed to test this
possibility.

\begin{figure}
  \plotone{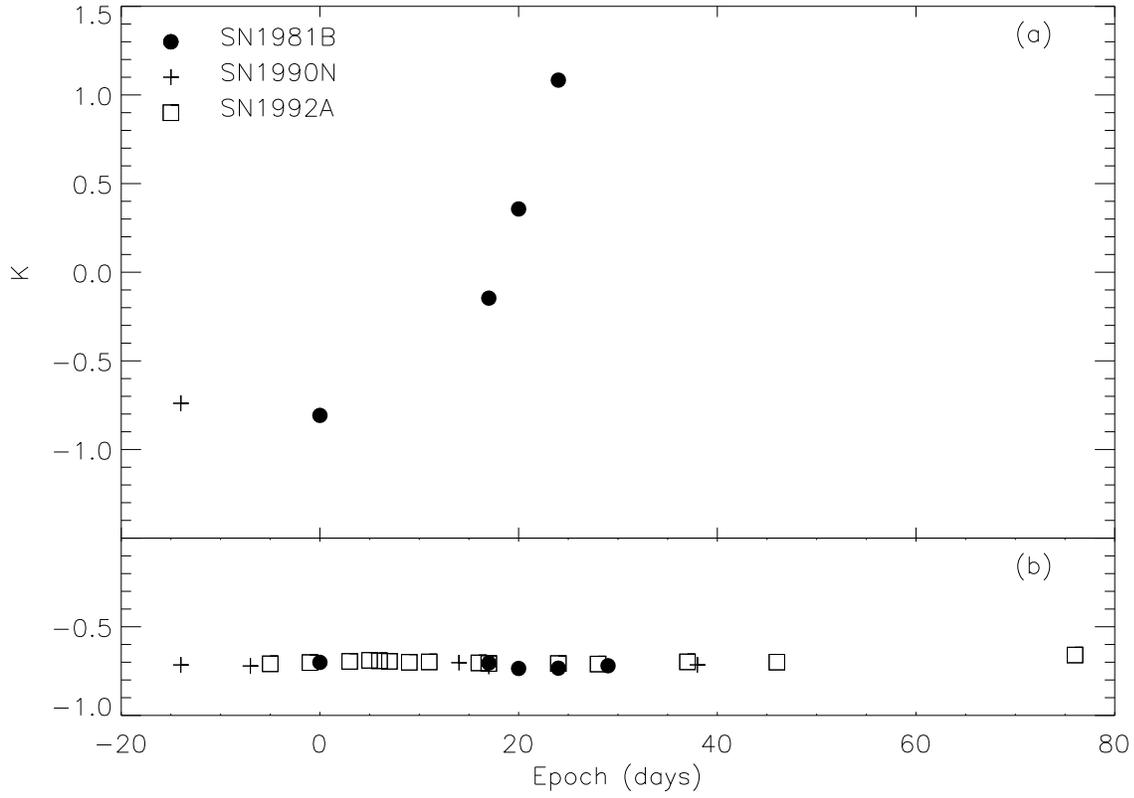}
  \caption[figurecaption]{(a)
  $K_{RR}(z=0.5)$ (or the $R$ band $K$ correction) as a function of epoch
  for SN1981B
  and SN1990N.  The available spectra of SN1992A does not have sufficient
  coverage to make such a calculation.
  (b) $K_{BR}(z=0.5)$ plotted on the same scale as (a) to show the
  relative ranges of the two corrections.}
  \label{krr5}
\end{figure}

In the above discussions, no conclusions could be made on
$K_{RR}$ error due
to sparseness of data.  However, error estimates can be made based on the
range of the correction.  
Recalling Equation~\ref{Kij},
we see that $K$ corrections are the sum of an overall offset due to the
different filter zeropoints plus a spectrally dependent term.
Smaller spectral terms will propagate smaller errors into the $K$ correction
than larger spectral terms.
In this analysis, the relative size of the spectral term is apparent
in the spread over time of the $K$ correction, since it is the only
time-dependent term.
Comparing the spread over time from
Figure~\ref{krr5}(a) with that from Figure~\ref{kvr5}, we see that
the spectral contribution in the standard single-filter $K$ correction,
$K_{RR}$, is almost twice that of $K_{VR}$, showing that the errors
in $K_{RR}$ are much larger than those of $K_{VR}$.
This point is demonstrated more dramatically
in Figure~\ref{krr5}(a,b)
where $K_{BR}$ and $K_{RR}$ are plotted on the same
scale.  It is clear that the scatter in $K_{BR}$ is much smaller than
would be expected for $K_{RR}$.

\begin{figure}
  \plotone{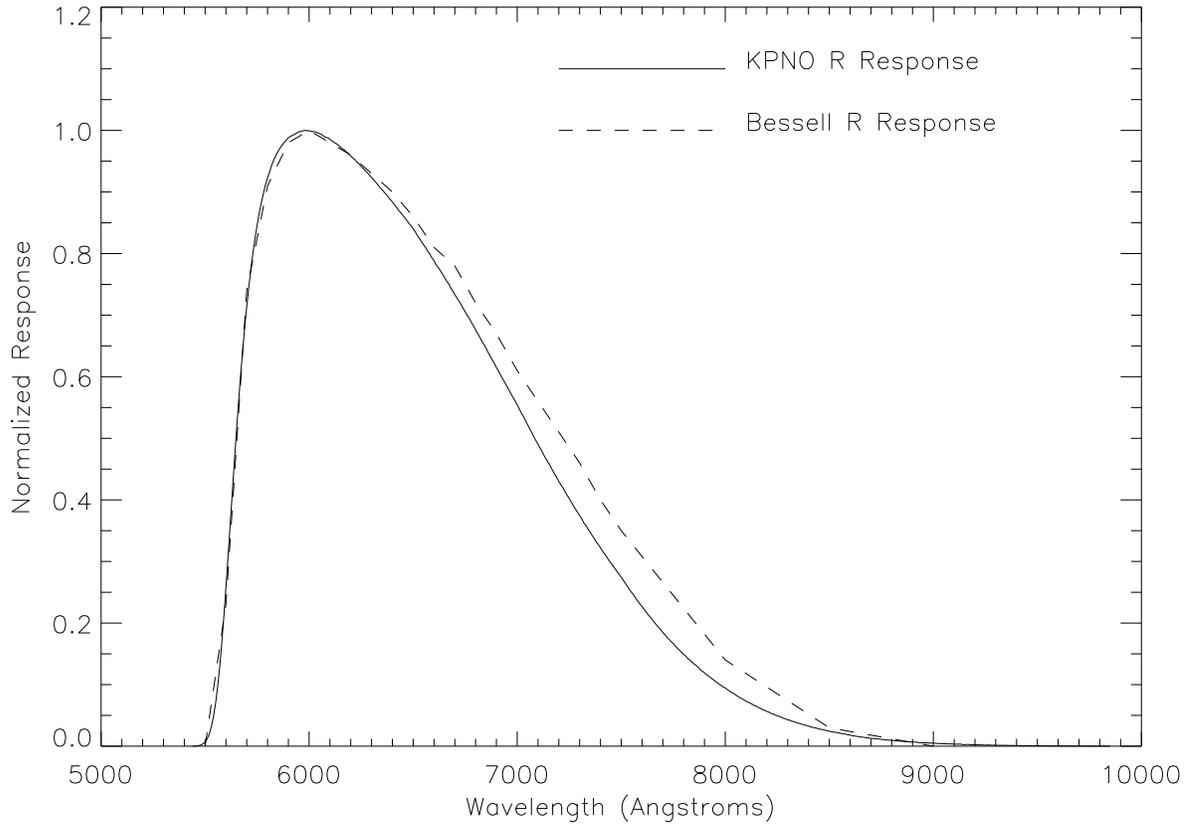}
  \caption[figurecaption]{
  Comparison between the Bessell representation of $R$ and our constructed
  response of the KPNO $R$ as described in the text,
  normalized at peak transmission.}
  \label{kpnofilter}
\end{figure}

\begin{figure}
  \plotone{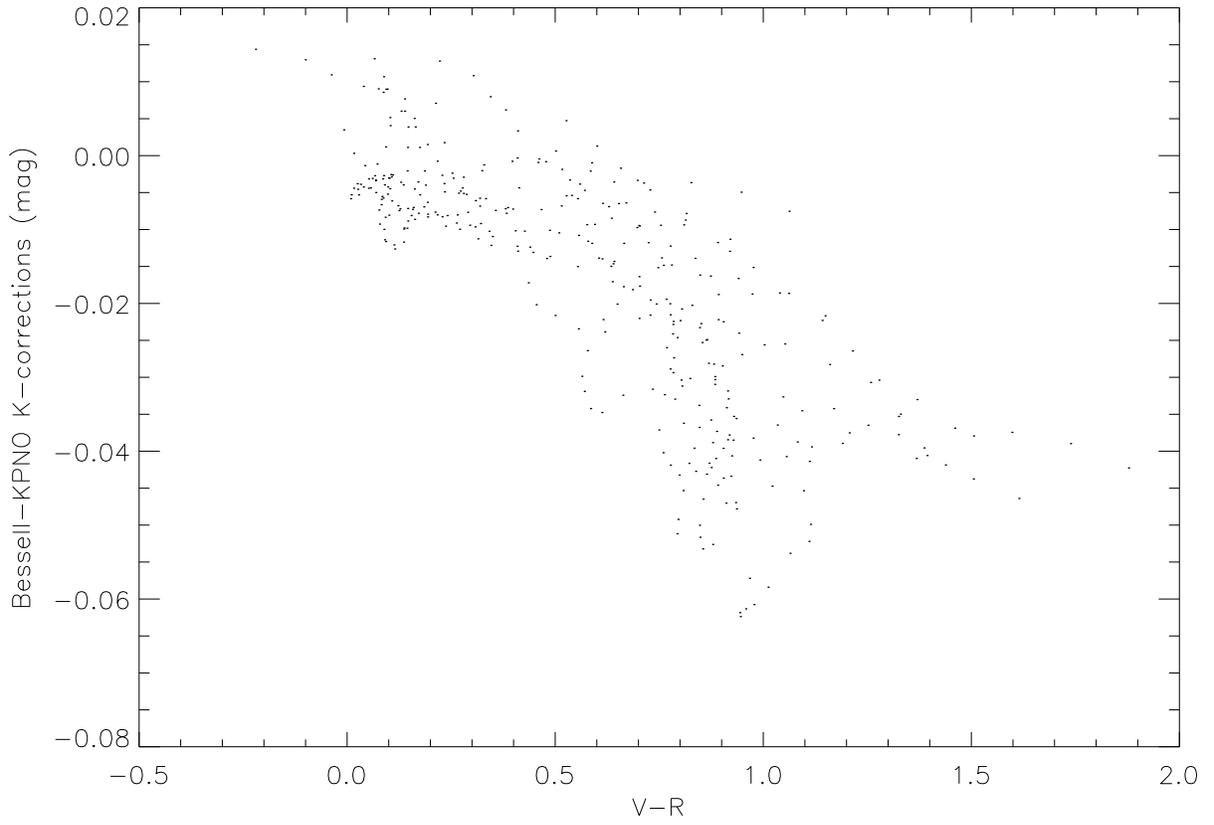}
  \caption[figurecaption]{
  $K_{BR}(Bessell) - K_{BR}(KPNO)$ as a function of observed color for all
  redshifts.}
  \label{BessminKPNOcol}
\end{figure}

We have examined potential
instrumental effects by performing $K_{xy}$ calculations
with effective Kitt Peak 4-m passbands based on the quantum efficiency of
the TK2B CCD camera and the KP1464 Harris B, KP1465  Harris V, and
KP1466 Harris R filters for the $B$,$V$, and $R$ filters respectively.
Observations of standard stars show that this effective passband
closely matches the Johnson-Cousins system, i.e. 
the color-correction term to transform from the instrumental to the
photometric system is negligible.
Comparison between these and the Bessell filter $K$ corrections show differences
at a level of 0.02 mag with standard deviation 0.02 mag.
We also see the expected correlation with the observed supernova
color; corrections for redder supernovae are more sensitive to the 
differences in the red tails of the Bessell and Kitt Peak
response functions.  (See Figure~\ref{kpnofilter}.)
The nature of this dependence can be seen in
Figure~\ref{BessminKPNOcol}.
A more detailed analysis will require a more complete spectral data set.

There is an additional error due to zeropoint
uncertainty for $K_{BR}$ and $K_{VR}$
mag which does not effect $K_{RR}$.  This is an advantage of the
single-filter $K$ correction, since zeropoints cancel if you compare
data in the same band.  As discussed earlier, the size of this error is
less than 0.01 mag.

Given these contributing sources of error, we can compare the overall
uncertainties for the generalized and standard 
$K$ corrections.  We begin with the case of $z=0.5$.
$K_{BR}$ and $K_{VR}$ have the same zeropoint uncertainty, however at 
this redshift $K_{BR}$ has smaller errors
than $K_{VR}$ from all other sources.  This includes
measurement error, instrumental systematics, and supernova-to-supernova
systematics.  Although $K_{RR}$ has no zeropoint error, it is otherwise expected
to have errors even larger than $K_{VR}$.
We emphasize that $K_{BR}$ has these advantages because at
$z=0.5$, $R(\lambda(1+z)) \approx
B(\lambda)$,  minimizing the spectral dependence on the $K$ correction
(Figure~\ref{filters}).

\begin{figure}
  \plotone{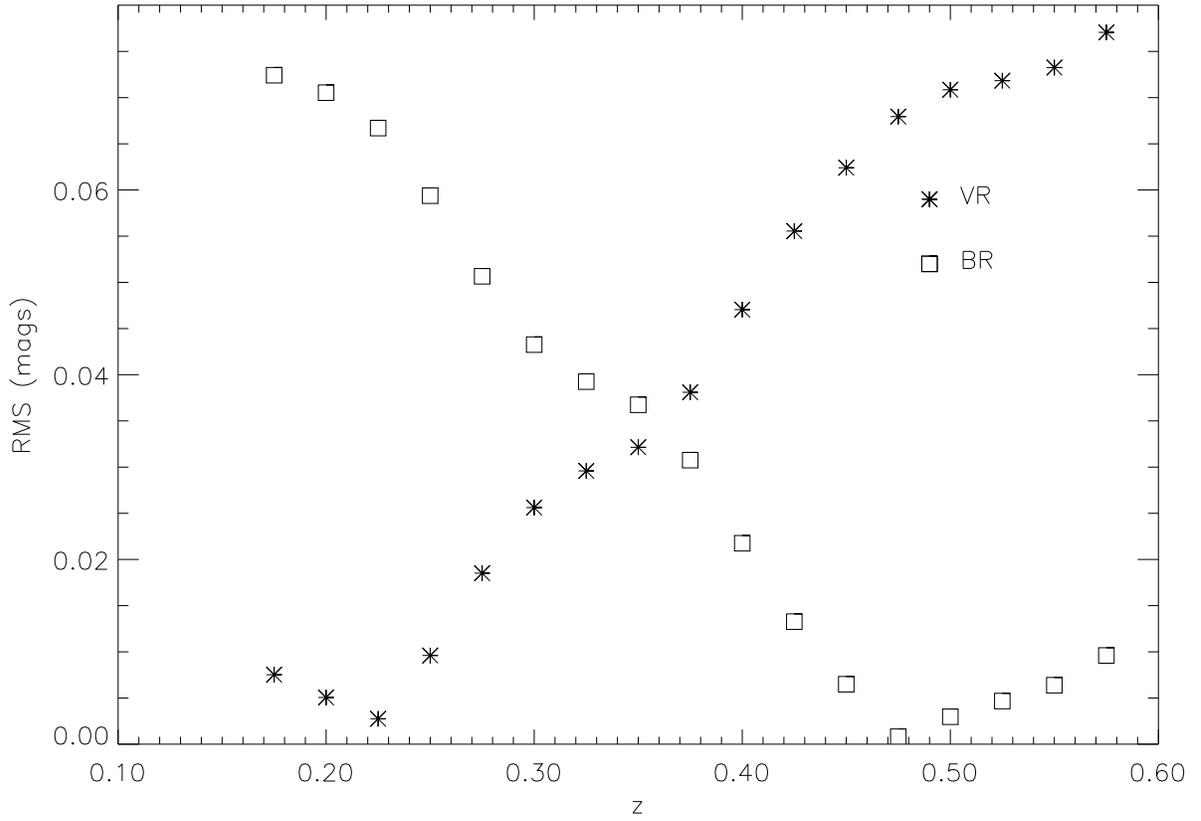}
  \caption[figurecaption]{
  The root-mean-square scatter of $K_{BR}$ (squares) and $K_{VR}$ (stars)
  for SN1992A at epochs -1 and 3, and
  SN1981B at epoch 0 as a function of redshift.}
  \label{stdevs}
\end{figure}

The case of $z=0.5$ is important as an extreme in which it is possible
to match filters.  
To illustrate the errors expected for other redshifts, Figure~\ref{stdevs}
also shows the calculated root-mean-square scatter
for the group of $K$ corrections near peak magnitude,
for SN1992A at epochs -1 and 3,
and SN1981B at epoch 0.
The root-mean-square scatter is minimized at redshifts
where the filters best match, and monotonically worsens as one
moves away from these redshifts.  Note that the error can be kept
below 0.04 mag by switching from the nearby $V$ photometry to the
nearby $B$ photometry when comparing supernovae at $z>0.36$.

\section{Conclusions}

We have considered a generalized $K$ correction as an alternative
to the single-band $K$ correction for relating local and high-redshift
supernova magnitudes.  Error sizes depend on the filter pair combination
chosen
and reflect the size of the term that accounts for the different spectral
regions observed in distant and local supernovae.  Minimizing this term
by matching filters to
observe the same region reduces error and can make
a generalized $K$ correction better than a single-band $K$ correction.
Matching filters also reduces the wavelength range needed to perform
$K$ corrections, making
more of the available data usable for better
temporal coverage and for studies in systematic differences in supernovae.
Generally, error estimates and optimal filter pair determinations at
any redshift can be
made using a
procedure similar to the one we have outlined for $z=0.5$. 
Roughly, we find that
for $z < 0.1$, $K_{RR}$ should be used, for $0.1 < z < 0.35$,
$K_{VR}$ should be used, and for $0.35< z < 0.7$, $K_{BR}$ should be used.
Objects at higher redshift may be better observed in the
$I$ band.

These general results for the preferred observation bands at a given redshift
to ``match filters''
will, of course, be true whatever the spectra observed.
However, further studies based on more supernova spectra will
improve the estimates of the generalized $K$ corrections, 
and help characterize the detailed dependence on
supernova-to-supernova variation.

We would like to thank B. Leibundgut, M. Bessell, D. Branch, and the
Calan/Tololo Search Group (M. Phillips, N.Suntzeff, J. Maza, and M. Hamuy)
for providing data and for helpful discussions.
This work was supported in part by the National Science Foundation
(ADT-88909616) and
U.~S. Department of Energy (DE-AC03-76SF000098).

\clearpage

\setcounter{table}{0}

\begin{table}
\renewcommand{\footnoterule}{\\}
\hspace{1.97in}T\scriptsize{ABLE} \normalsize 1

\scriptsize
\hspace{1.57in}Selected Spectra of SNe Ia\\

\begin{tabular}{c c c c c}
\tableline \tableline
SN & Epoch \tablenotemark{a} & UT Date & Observatory/Tel & Observer(s)\\ \tableline
1990N & -14 & 1990 Jun. 26.17 & MMTO/MMT & Foltz \\
1990N & -7 & 1990 Jul. 02.99 & CTIO/1.5-m & Phillips \\
1992A & -5 & 1992 Jan. 14.11 & CTIO/1.0-m & Winge \\
1981B & -1(a) & 1981 Mar. 6 & McDonald/2.7-m & See Branch et al. (1983)\\
1981B & -1(b) & 1981 Mar. 6 & McDonald/2.7-m & See Branch et al. (1983)\\
1992A & -1 & 1992 Jan. 18.13 & CTIO/1.0-m & Winge \\
1981B & (0)\tablenotemark{b} & 1981 Mar. 7 & McDonald/2.7-m & See Branch et al. (1983)\\
1981B & 0 & 1981 Mar. 7 & McDonald/2.7-m & See Branch et al. (1983)\\
1992A & +3 & 1992 Jan. 22.04 & CTIO/1.0-m & Winge \\
1992A & (+5)\tablenotemark{b} & 1992 Jan. 24 & HST & SINS\tablenotemark{c} \\
1992A & +6 & 1992 Jan. 25.04 & CTIO/1.5-m & Smith/Winkler \\
1992A & +7 & 1992 Jan. 26.04 & CTIO/1.0-m & Winge \\
1990N & +7 & 1990 Jul. 17 & Lick/3.0-m & Shields/Filippenko \\
1992A & +9 & 1992 Jan. 28.04 & CTIO/1.5-m & Hamuy/Williams \\
1992A & +11 & 1992 Jan. 30.04 & CTIO/1.0-m & Winge \\
1990N & +14 & 1990 Jul. 23.98 & CTIO/4.0-m & Phillips/Baldwin \\
1992A & +16 & 1992 Feb. 04.04 & CTIO/1.0-m & Winge \\
1981B & +17 & 1981 Mar. 24 & McDonald/2.7-m & See Branch et al. (1983)\\
1992A & +17 & 1992 Feb. 05.04 & CTIO/4.0-m & Hamuy \\
1990N & +17 & 1990 Jul. 27.16 & MMTO/MMT & Huchra \\
1981B & +20 & 1981 Mar. 27 & McDonald/2.7-m & See Branch et al. (1983)\\
1981B & +24 & 1981 Mar. 31 & McDonald/2.7-m & See Branch et al. (1983)\\
1992A & +24 & 1992 Feb. 12.03 & CTIO/1.0-m & Winge \\
1992A & +28 & 1992 Feb. 16.02 & CTIO/1.0-m & Winge \\
1981B & +29 & 1981 Apr. 5 & McDonald/2.7-m & See Branch et al. (1983)\\
1992A & +37 & 1992 Feb. 25.01 & CTIO/1.0-m & Winge \\
1990N & +38 & 1990 Aug. 16.98 & CTIO/1.5-m & Phillips \\
1992A & +46 & 1992 Mar. 05.02 & CTIO/1.5-m & Phillips \\
1981B & +49 & 1981 Apr. 25 & McDonald/2.1-m & See Branch et al. (1983)\\
1981B & +58 & 1981 May. 4 & McDonald/2.7-m & See Branch et al. (1983)\\
1992A & +76 & 1992 Apr. 04.02 & CTIO/4.0-m & Hamuy/Maza \\
\tableline \tableline
\tablenotetext{a}{Epoch relative to $B$ maximum light.}
\tablenotetext{b}{See text for discussion.}
\tablenotetext{c}{Kirsher et al. (1988)\nocite{ki:sins}}
\label{spectra}
\end{tabular}
\end{table}

\newpage
\normalsize

\begin{figure}
  \plotfiddle{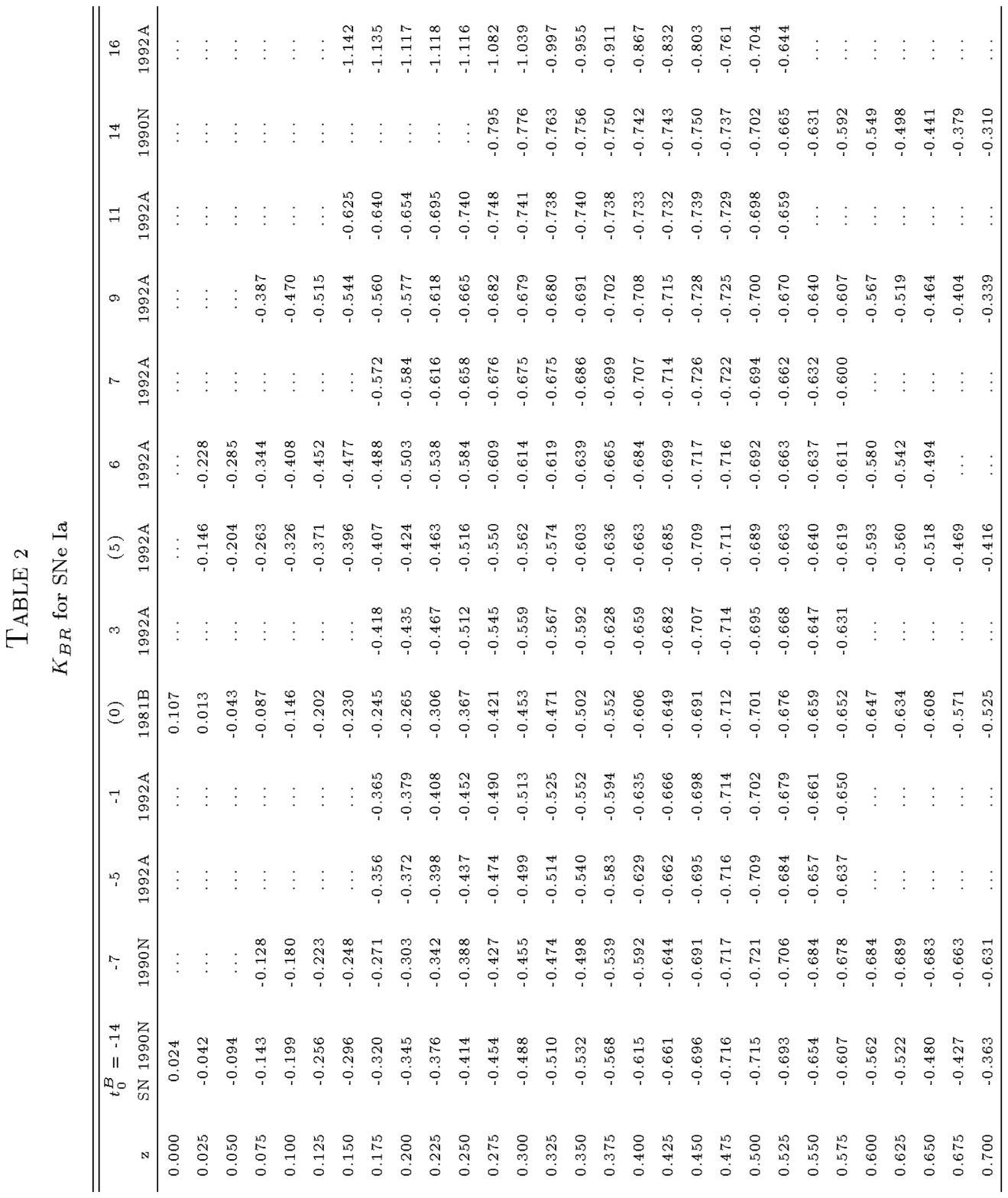}{12in}{0}{115}{115}{-360}{120}
  \label{br1table}
\end{figure}

\begin{figure}
  \plotfiddle{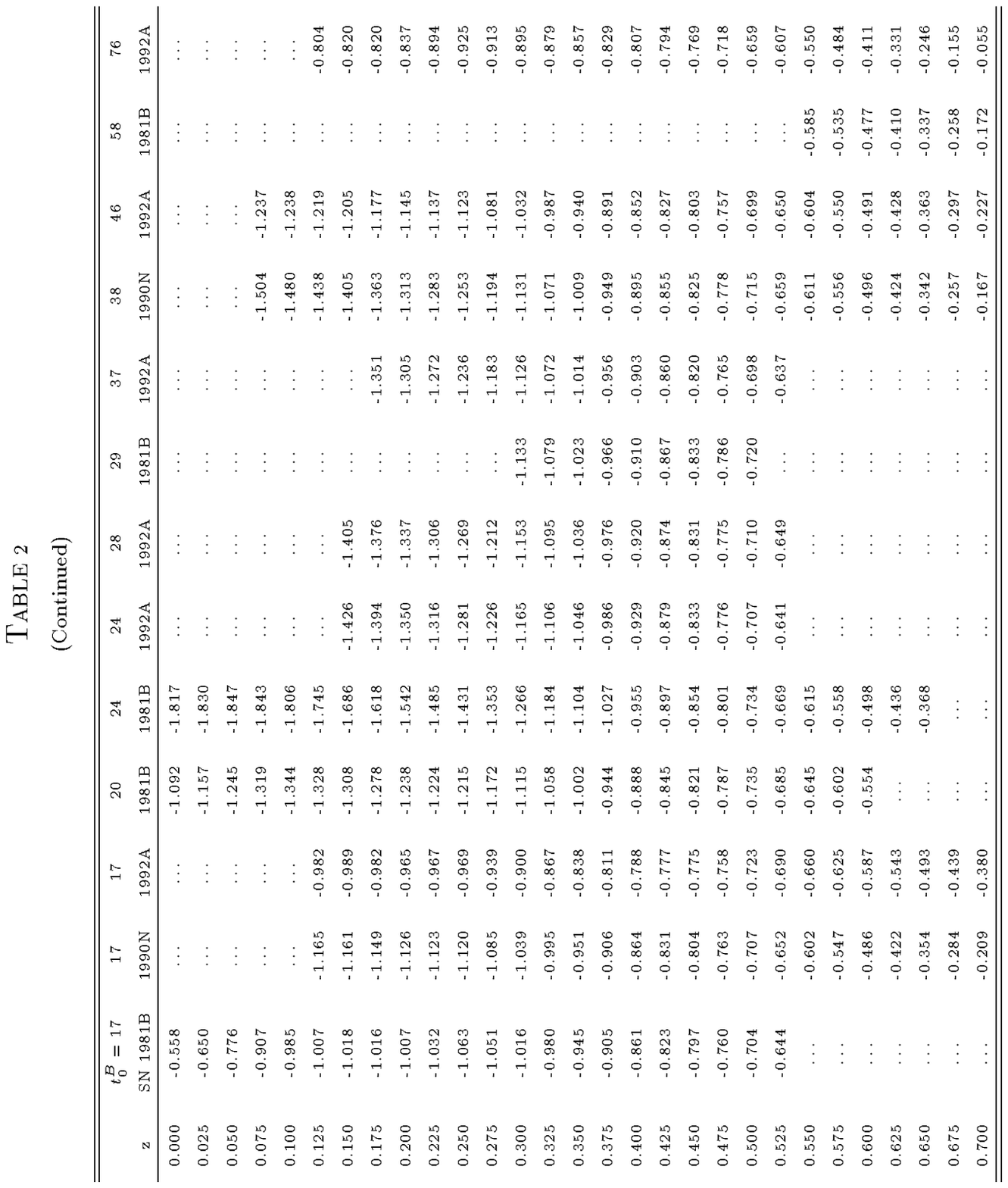}{12in}{0}{115}{115}{-360}{120}
  \label{br2table}
\end{figure}

\begin{figure}
  \plotfiddle{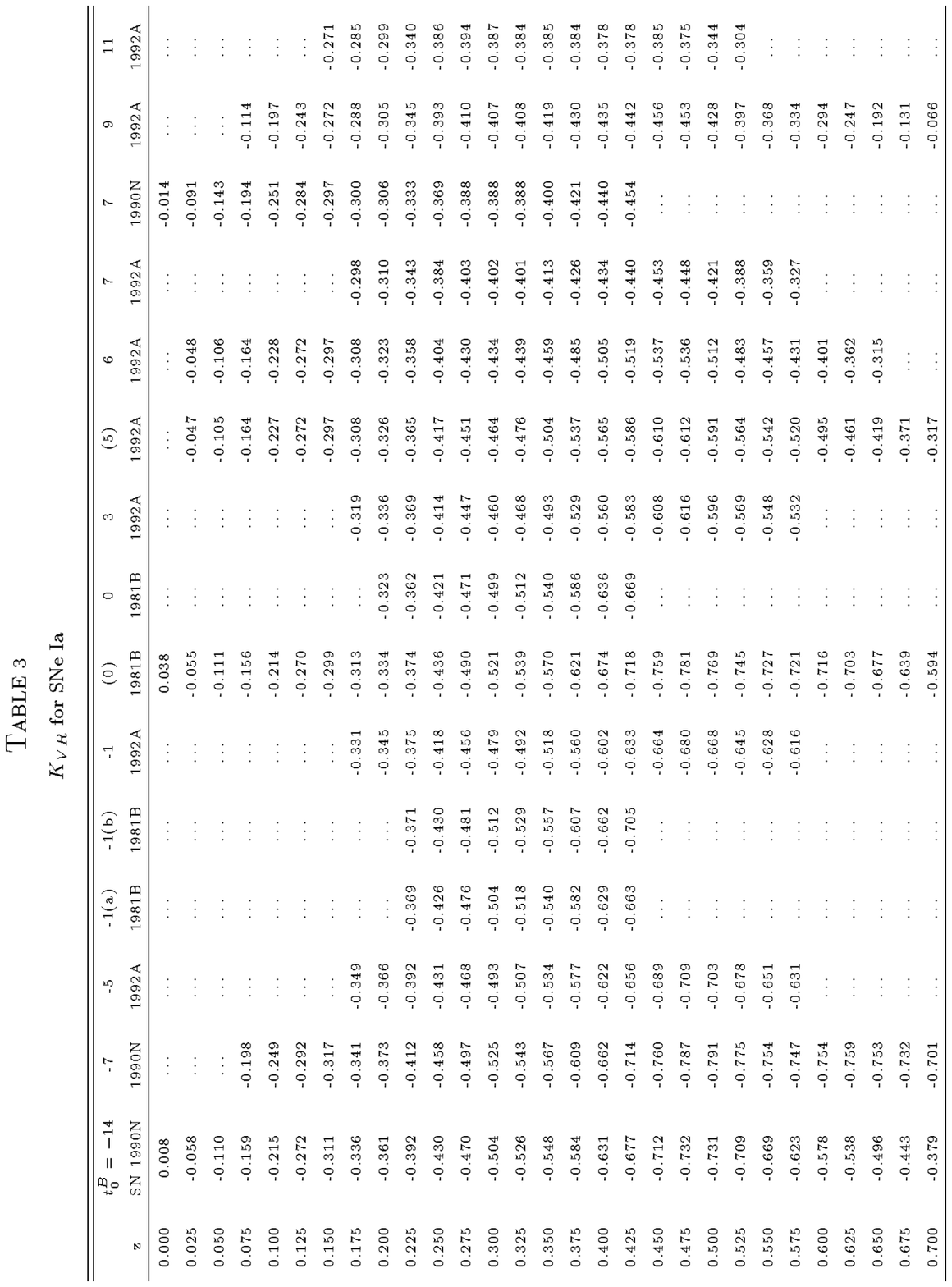}{12in}{0}{115}{115}{-360}{110}
  \label{vr1table}
\end{figure}

\begin{figure}
  \plotfiddle{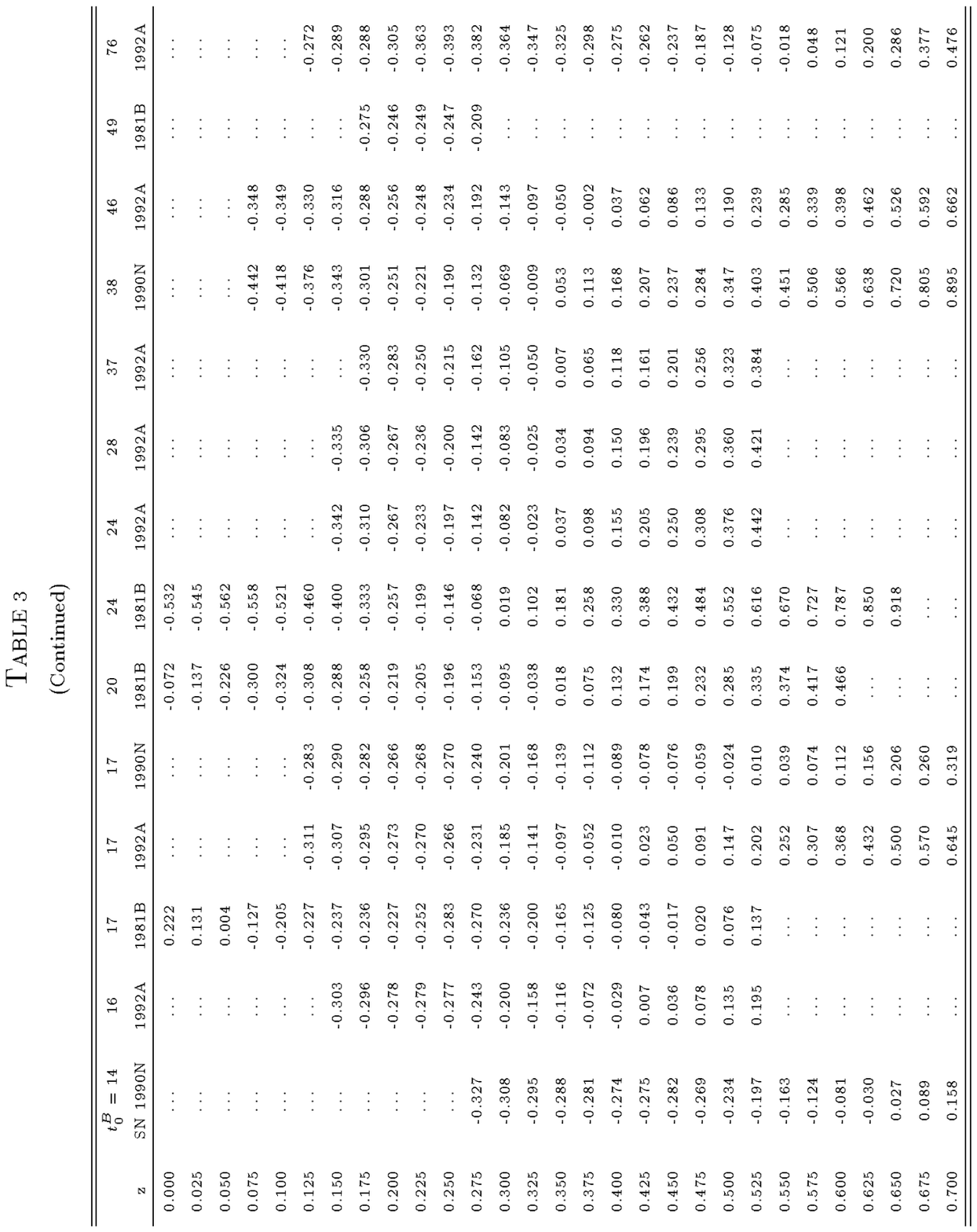}{12in}{0}{115}{115}{-360}{120}
  \label{vr2table}
\end{figure}

\begin{figure}
  \plotfiddle{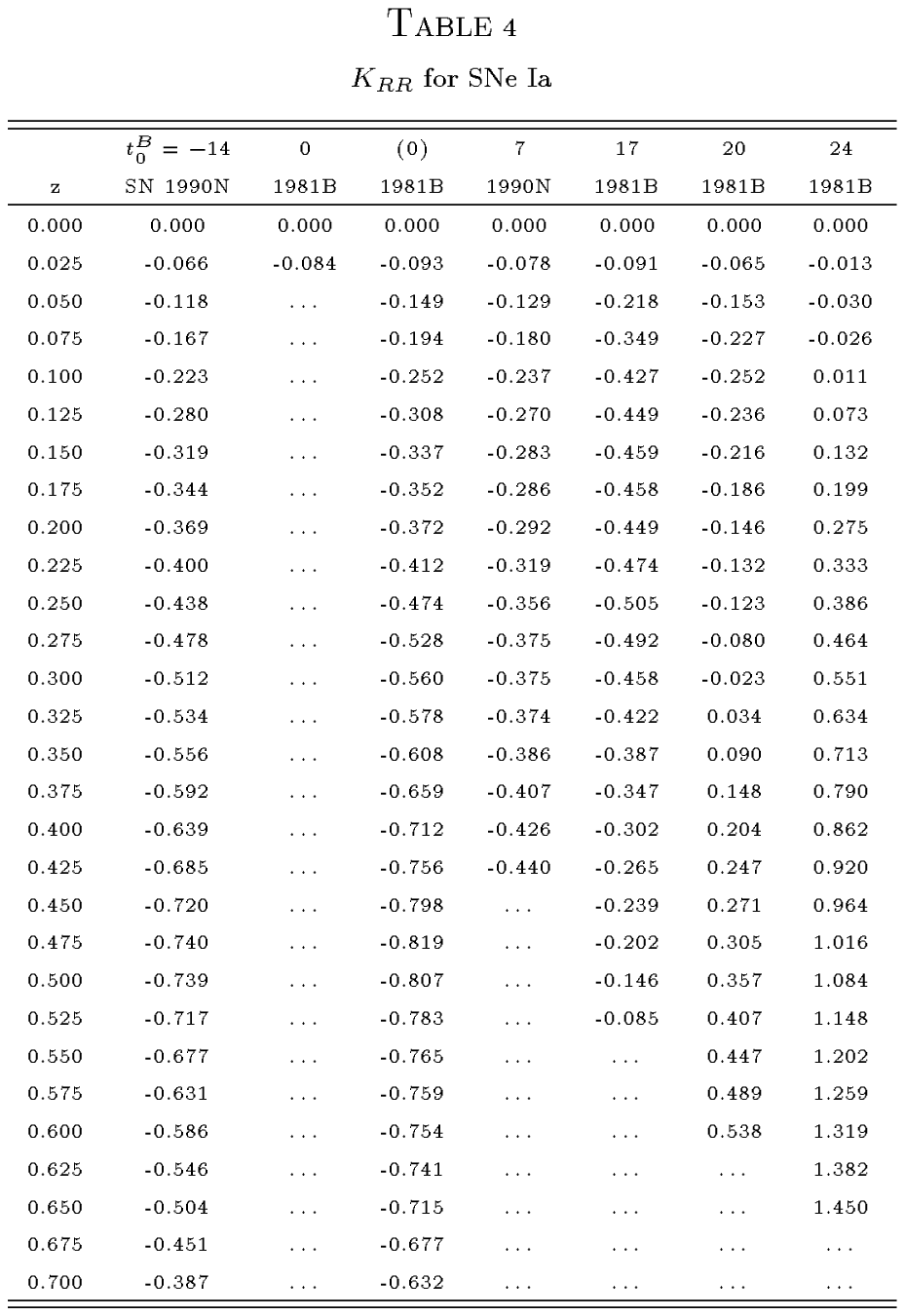}{12in}{0}{100}{100}{-288}{120}
  \label{rrtable}
\end{figure}
\newpage

\end{document}